\documentclass[runningheads]{llncs}
\usepackage{graphicx}
\usepackage{amsmath, amsfonts}
\usepackage{algorithm2e}
\usepackage{hyperref}
 
\usepackage{xcolor}

\usepackage{todonotes}

\begin{document}

\title{Operating with Quantum Integers:\\an Efficient `Multiples of' Oracle}

\titlerunning{`Multiples of' oracle}

%

\author{Javier Sanchez-Rivero\inst{1} \and
Daniel Talaván\inst{1} \and
Jose Garcia-Alonso\inst{2} \and
Antonio Ruiz-Cortés \inst{3} \and
{Juan Manuel} Murillo \inst{1,2}}
\authorrunning{J. Sanchez-Rivero, D. Talaván et al.}
%
\institute{COMPUTAEX  \and
University of Extremadura \and
Universidad de Sevilla}

\maketitle              
%


\begin{abstract}

Quantum algorithms are a very promising field. However, creating and manipulating these kind of algorithms is a very complex task, specially for software engineers used to work at higher abstraction levels. The work presented here is part of a broader research focused on providing operations of a higher abstraction level to manipulate integers codified as a superposition. These operations are designed to be composable and efficient, so quantum software developers can reuse them to create more complex solutions. Specifically, in this paper we present a `multiples of' operation. To validate this operation we show several examples of quantum circuits and their simulations, including its composition possibilities. A theoretical analysis proves that both the complexity of the required classical calculations and the depth of the circuit scale linearly with the number of qubits. Hence, the `multiples of' oracle is efficient in terms of complexity and depth. Finally, an empirical study of the circuit depth is conducted to further reinforce the theoretical analysis.

\keywords{Quantum computing  \and Amplitude Amplification \and Oracle \and Multiples \and Qiskit}
\end{abstract}

\section{Introduction} \label{sec:intro}

Quantum computing \cite{national2019quantum} uses quantum mechanics to perform computations in a different manner than classical computing \cite{nielsen2002quantum}. Nowadays, quantum computers are in the NISQ (Noisy Intermediate-Scale Quantum) Era \cite{nisq}, which means their practical use is still limited by errors and the low number of qubits (quantum bits). However, the recent developments on quantum devices has allowed researchers to start testing on real quantum hardware the theoretical work on quantum algorithms, which has been a very active field for decades \cite{history}.

Quantum algorithms are useful when they can solve certain problems faster than any known classical algorithm \cite{montanaro2016quantum}. This speedup is measured in terms of asymptotic scaling of complexity \cite{bigOnotation}. The work presented here is part of ongoing research aimed at providing programmers with operations on quantum states at a higher level of abstraction than the base quantum gates. More specifically, our research has begun with the goal of providing operations on a superposition quantum state that encodes integers with size determined by the number of qubits in the state. These operations are not only useful for manipulating a quantum state encoding integers, they are also more efficient than the same operations in the classical domain. In addition, the quantum circuits that implement these operations are optimised in depth, as well as the number of qubits (ancilla and non-ancilla) they use.

In particular, this paper presents an operation that computes multiples. Thus, given an integer and a quantum state that encodes integers, the operation phase-tags the configurations of the quantum state that correspond to multiples of the given integer. While the complexity of this operation in the classical domain is $\mathcal{O}(2^n)$, the complexity of the operation presented here is $\mathcal{O}(n)$ where $n$ is the number of bits codifying the maximum size of the wanted multiples. This is a logarithmic scaling in the total number of states, which provides an exponential speedup with respect to classical calculations.

As mentioned before, in our research this 'multiples of' is part of a larger set of quantum operations on integers \cite{sanchez2023automatic,sanchezrivero2023initial}. An important feature that we want the whole set of operations to preserve is composability. Thus, all the resources of this set are composable with each other and, for example, the `multiples of' can be composed with a `less than' operation to obtain the multiples of a given integer $a$ less than another given integer $b$. In particular, the operation `multiples of' can also be composed with itself to, for example, mark in a quantum state the multiples of an integer $a$ and then, in the resulting state, mark the multiples of another integer $b$. Preserving composability offers the possibility to reuse such operations to build more complex operations with a higher level of abstraction. Achieving the best quality attributes in each operation is important because their compositions will inherit those attributes \cite{Zhao,klappenecker2003quantum}.

This paper is organized as follows. In section \ref{sec:background} we provide the necessary background for this work. Next, the description of the 'multiples of' operation is presented in section \ref{sec:implementation_of_oracle}. The operation takes the shape of a quantum oracle and the section details both the idea inspiring the oracle and the quantum circuit exact implementation. In section \ref{sec:simulations} some examples of circuits and simulations are shown to prove the functionality of the oracle. Then, in section \ref{sec:complexity}, the complexity of classical calculations as well as the quantum circuit is discussed. Section \ref{sec:composability} shows composability and further uses of the oracle. Finally, in section \ref{sec:conclusions}, the conclusions and future work are explored.

\section{Background} \label{sec:background}

Oracles have been identified as a recurring pattern in quantum software \cite{Leymann_QuantumAlgorithms,Zhao}. Following this trend, the work presented here is built as an oracle for Grover's algorithm \cite{grover1}.

This algorithm searches for one quantum state in an unordered database faster than any known classical algorithm. Its generalization is called Amplitude Amplification \cite{Grover_1998,AA} and allows to search for multiple values. This algorithm works by applying two quantum operations: an oracle, which marks with a $\pi$-phase the desired quantum states, and a diffuser, which tries to amplify the amplitude of those marked states. Often, to reach a satisfactory amplification, it is needed to repeat the pair oracle-diffuser several times. This pair oracle-diffuser is usually called the Grover iterator.

The `multiples of' oracle is built reusing two pieces of existing quantum software, the linear multi-controlled gate \cite{multicontrol2022} and the modulo addition \cite{draper,Shor23}. 

In \cite{multicontrol2022} the authors present an efficient implementation of a multi-controlled gate whose depth scales linearly with the number of qubits and thus avoids the polynomial growth of previous implementations. Furthermore, it does not require the use of ancilla qubits. The linear multi-control gate outperforms Qiskit\cite{Qiskit} implementation from five qubits onwards, which supposes a clear improvement for any meaningful use. Because of these quality attributes we chose to reuse it in this work.

The modulo addition operation $a+b$ mod $k$ is defined as the remainder of dividing $a+b$ by $k$. In this work, this operation will only be performed with values $a,b<k$. Hence, in our case, the modulo addition can be written as:
\begin{equation*}
a+b \text{ mod } k =
\left\{
\begin{array}{lc}
    a+b & \ \text{ if } \ a+b < k \\
   a+b-k  & \ \text{ if } \ a+b \geq k 
\end{array}
\right. \quad a,b<k
\end{equation*}

For this modulo addition we use the implementation presented in \cite{Shor23}. It is heavily based on Draper's algorithm \cite{draper} for quantum addition. This method uses the quantum Fourier transform \cite{QFTcoppersmith} and hence is done on the frequency domain. It allows the addition of an integer to a quantum superposed state without the need to encode the integer in a quantum register. This reduces the number of necessary qubits. The depth of this operation is linear on the number of qubits, as it is a composition of linear-depth primitives.

Once the addition gate is built, the modulo addition conducts the following operations: adds a classical value $a<k$ to a quantum state holding the classical value $b<k$, and then it substracts $k$ if $a+b\geq k$. This methods requires two ancilla qubits to perform the operation, one for the overflowing of the sum, and another one for checking whether it is needed to substract $k$ or not.

This paper showcases how by carefully composing existing pieces of quantum software a new non-trivial software can be obtained. The ideas hereby described are the ones which allow to build the oracle `multiples of', presented in detail in the next section.

\section{Implementation of the `Multiples of' Oracle} \label{sec:implementation_of_oracle}

In this section we provide a description of the oracle. It comprises two differentiated parts, the first one is the mathematical ideas inspiring the oracle, where basic modulo theory shows a condition for identifying multiples. The second part describes the quantum circuit of the oracle, how the multiples are given a $\pi$-phase, and how to adapt the oracle for a full Amplitude Amplification implementation.

\subsection{Mathematical Properties Inspiring the Oracle} \label{subsec:theory}
A number $M\in \mathbb{N}$ is multiple of another number $k$ if the remainder of the division is 0, formally expressed as $M\equiv 0$ mod $k$. If $M$ is not a multiple of $k$, then \mbox{$M \not\equiv 0$ mod $k$}.

The number $M$ can be be expressed in binary form, also known as binary decomposition:
\begin{equation}
M = a_m \cdot 2^m + a_{m-1}\cdot 2^{m-1} + \ldots + a_1\cdot 2^1 + a_0\cdot 2^0 = \sum_{i=0}^n a_i\cdot 2^i
\end{equation}

where $a_i \in \{0, 1\}$ and $m = \lceil \log_2 M \rceil$.\\

Let $r_i$, with $0\leq r_i < k$, be the remainder of $2^i$ when divided by $k$, formally:
\begin{equation}
    2^i \equiv r_i \text{ mod } k
\end{equation}
Then by the properties of the ring of remainders $\mathbb{Z}_k$ it can be noticed that:
\begin{equation}
    M \equiv \sum_{i=0}^m a_i\cdot 2^i \equiv \sum_{i=0}^m a_i\cdot r_i \text{ mod } k
\end{equation}

Hence, $M$ is a multiple of $k$ if the sum of the remainders of the powers of 2 modulo $k$ of its binary decomposition is equivalent to 0, formally:

\begin{equation}\label{eq:remainder-M}
    M \equiv 0 \text{ mod } k \ \Longleftrightarrow \ \sum_{i=0}^m a_i\cdot r_i \equiv 0 \text{ mod } k
\end{equation}

Therefore, the `multiples of' oracles is built in two parts, first adding the remainders of the powers of two, and then givin a $\pi$-phase to those which are 0, thus the multiples.

\subsection{Algorithm for the `Multiples of' Oracle}\label{sub:algorithm-oracle}
This subsection provides a detailed explanation of the implementation of the `multiples of' oracle.

Let $k\in \mathbb{N}$ be the number whose multiples want to be calculated. The quantum circuit for the `multiples of $k$' oracle consists of three registers of qubits.

The first is the input register, which holds the quantum states in which the multiples will be searched. It is formed by $n$ qubits and the $i$-th qubit of this register is denoted $q_i$. The number $n$ is an input parameter and does not depend on any other value. Thus, the numbers in which the multiples will be searched range from 0 to $N-1$, where $N=2^n$.

The second register holds the remainder of the numbers. At most, the remainder of dividing by $k$ is $k-1$, hence the required number of qubits for this register is $n_k = \lceil \log_2 (k-1) \rceil$. The $i$-th qubit of this register is denoted $rq_i$.

Finally, an ancilla register with two qubits is needed to perform the modulo addition introduced in the section \ref{sec:background}, as described in detail in \cite{Shor23}. These qubits are denoted ancilla$_0$ and ancilla$_1$. Both the registers and the ancilla registers are initialized to state $|0\rangle$. 

The algorithm for building the circuit can be found in \ref{algorithm}. It follows an explanation which describes it thoroughly.

\SetKwComment{Comment}{/* }{ */}
\newcommand{\forcond}{$i=0$ \KwTo $n-1$}

\begin{algorithm}[h]
\caption{Algorithm for building the `multiples of' oracle}\label{algorithm}
\KwData{Number of qubits $n$ and a natural number $k$}
\KwResult{Quantum Circuit which gives a $\pi$-phase to all states representing binary forms of natural numbers multiples of $k$}
Calculate $r_i\equiv 2^i$ mod $k$ for $i \in [0, n-1]$ \Comment*[r]{see \hyperref[app:A]{Appendix A}}
$n_k \gets \lceil\log_2(k-1)\rceil$\;
input\_register ($q$) $\gets QuantumRegister(n)$\;
remainder\_register ($rq$) $\gets QuantumRegister(n_k)$\;
ancilla\_register $\gets QuantumRegister(2)$\;
$n_{total} \gets n + n_k + 2$\;
$circ \gets QuantumCircuit(n_{total})$\;
\For{\forcond}{
    $circ^{+} = ModuloAddition(r_i, n_k + 2)$\;
    append $circ^{+}$ to $circ$:\\
    $\quad$- Target: remainder\_register and ancilla\_register\;
    $\quad$- Control: $q_i$\;
}

\For{$j=0$ \KwTo $n_k-1$}{
    $X$ gate to $rq_j$\;
}
$CZ^{\otimes (n_k)}$ gate to qubits $rq_{0},\,\ldots,\,rq_{n_k-1}$\;
\For{$j=0$ \KwTo $n_k-1$}{
    $X$ gate to $rq_j$\;
}

\For{\forcond}{
    $circ^{-} = ModuloSubstraction(r_i, n_k + 2)$\;
    append $circ^{-}$ to $circ$:\\
    $\quad$- Target: remainder\_register and ancilla\_register\;
    $\quad$- Control: $q_i$\;
}

\end{algorithm}

\begin{figure}
    \centering
    \includegraphics[width = 0.7\textwidth]{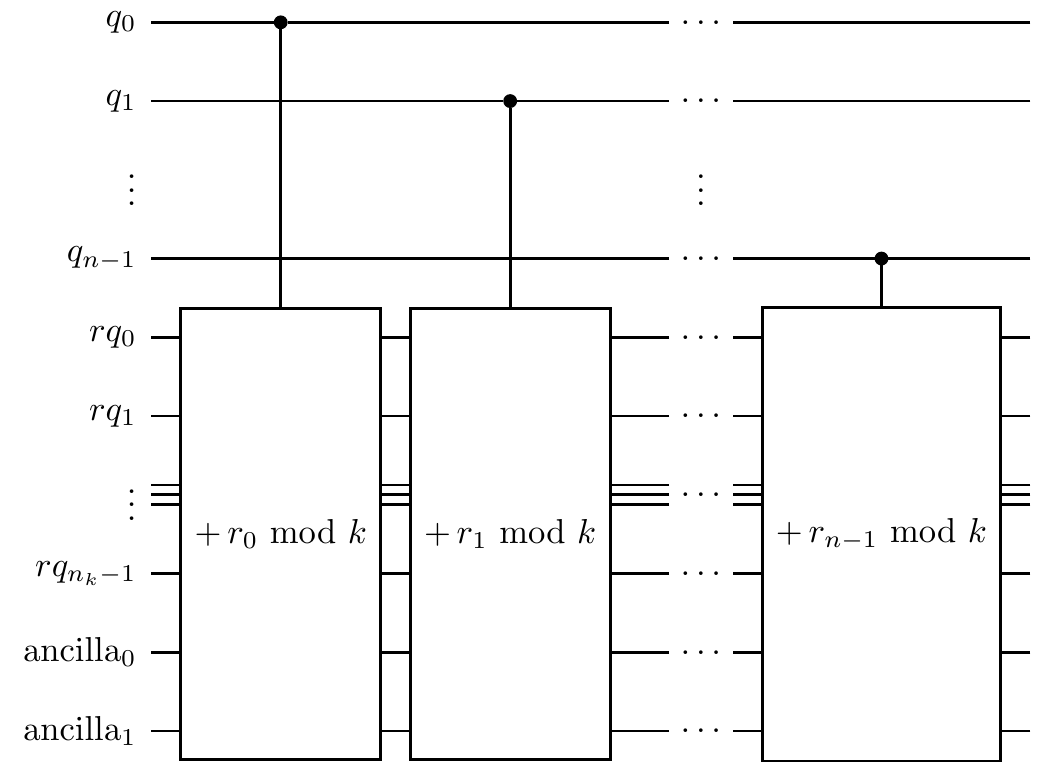}
    \caption{Modulo addition of the remainders of the powers of 2.}
    \label{fig:circuit-sum-remainders}
\end{figure}

The remainders of each power of 2, $r_i\equiv 2^i$ mod $k$, are added modulo $k$ to the remainders register, where the addition is controlled by the input qubit $q_i$. As the remainders register is initialized as $|0\rangle$ and $r_i < k \ \forall i$, the result of each modulo addition will never be larger than $k$, as shown in the definition of this operation in section \ref{sec:background}. Figure \ref{fig:circuit-sum-remainders} shows the general case of this implementation. This image and all the others showing circuits have been done with the \textit{quantikz} package \cite{quantikz}.

After applying the modulo additions, the ancilla register is always at state $|00\rangle$ \cite{Shor23}. The remainders register holds states from $|0\rangle$ up to $|k-1\rangle$. From equation \ref{eq:remainder-M}, it can be seen that the multiples of $k$ are those states of the form:

\begin{equation}
    |rq_{n_k-1}\ldots rq_1 \, rq_0 \, q_{n-1} \ldots q_1\,q_0\rangle=|\underbrace{0\ldots 0 \, 0}_{rq \text{ register}} \, \underbrace{q_{n-1} \ldots q_1\,q_0}_{\text{input register}}\rangle
\end{equation}

\begin{figure}
    \centering
    \includegraphics[width = 0.9\textwidth]{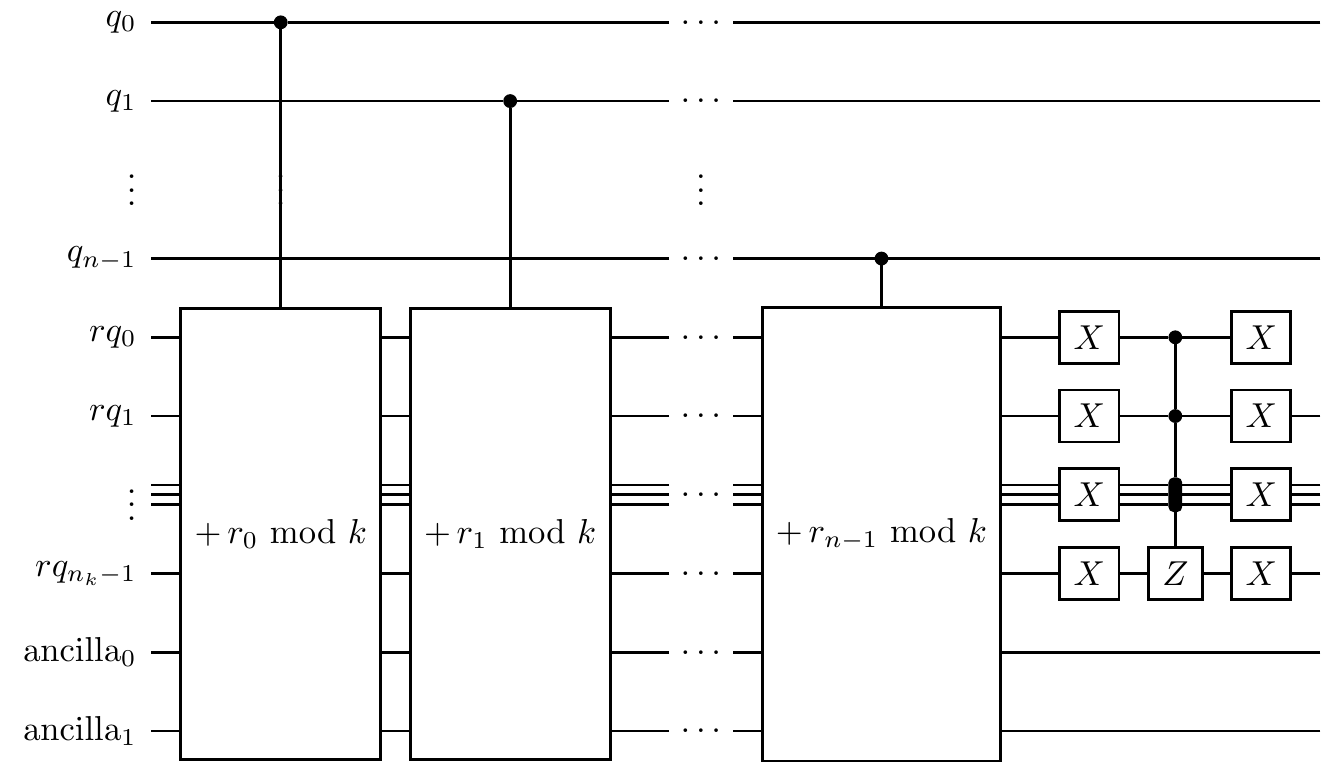}
    \caption{Oracle that marks multiples of $k$.}
    \label{fig:circuit-marking-multiples}
\end{figure}

Hence, these states and only these ones will be given a $\pi$-phase by means of the gate \ref{eq:marking-multiples}:
\begin{equation}\label{eq:marking-multiples}
    X^{\otimes n_k}\cdot CZ^{n_k} \cdot X^{\otimes n_k}
\end{equation}
where $CZ^{n_k}$ is a multi-controlled $Z$-gate whose target is qubit $rq_{n_k-1}$ and controlled by qubits $rq_0, \ldots, rq_{n_k-2}$. This gate is built using the linear multi-controlled $Z$-gate introduced in section \ref{sec:background}. Figure \ref{fig:circuit-marking-multiples} shows how this part of the circuit is built.

\begin{figure}
    \centering
    \includegraphics[width = \textwidth]{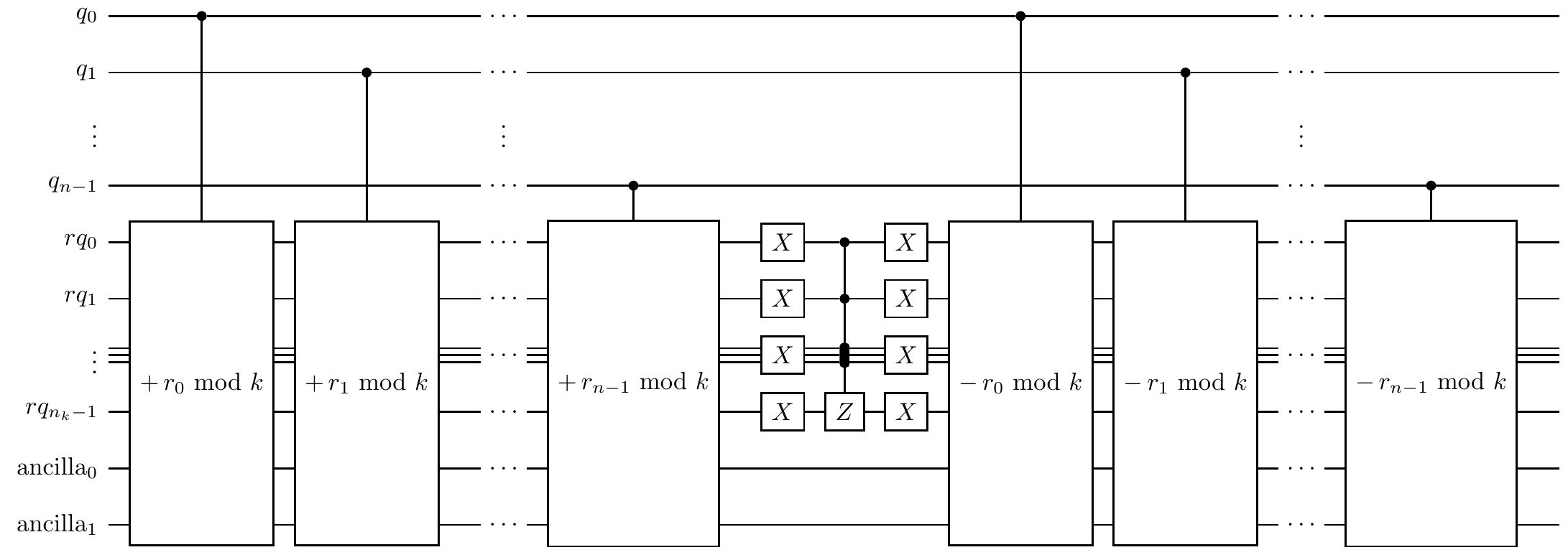}
    \caption{Oracle that marks multiples of $k$ and returns all auxiliary qubits to state $|0\rangle$.}
    \label{fig:circuit-full-oracle-multiples}
\end{figure}

Finally, in order to apply the diffuser, it is required to return auxiliary qubits to their initial states, that is, to perform an uncomputation on these register\cite{nielsen2002quantum}. As the multiples are already given a $\pi$-phase, if the modulo additions of the remainders are uncomputed, the states would keep the phase. The circuit would be as in figure \ref{fig:circuit-full-oracle-multiples}.

\begin{figure}
    \centering
    \includegraphics[width=\textwidth]{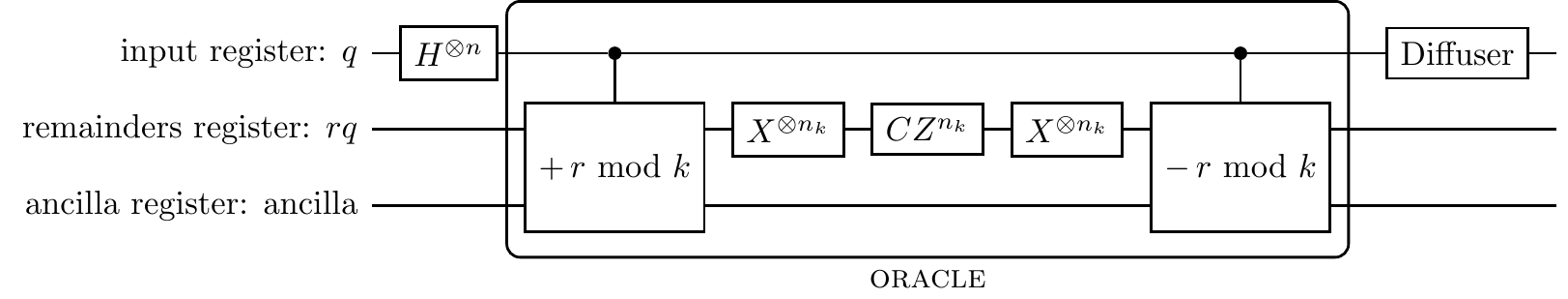}
    \caption{Implementation of `multiples of' oracle plus diffuser with full superposition as input.}
    \label{fig:circuit-oracle-plus-diffusers}
\end{figure}

Finally, the diffuser can be applied to the input register (the rest of the registers, remainders and ancilla, are in the state $|0\rangle$). The complete implementation is shown in figure \ref{fig:circuit-oracle-plus-diffusers}.

A documentation for this `multiples of' oracle following the guidelines proposed in \cite{sanchezrivero2023initial} can be found in the following \hyperref[sec:repository]{repository}.

\section{Simulations and Results}\label{sec:simulations}
In this section we show some examples of the oracle. Different values for $k$, the number whose multiples are calculated, and $n$, the number of input qubits\footnote{We codify all our integers as quantum states in these $n$ qubits, hence the multiples are calculated up to $N=2^n$ integers.}, are chosen to showcase the functionality of the `multiples of' oracle. We display the full circuit for the multiples of 3 oracle with 4 qubits input as well as its simulation. We also show the full circuit and the simulation of multiples of 5 with 6 qubits. We have chosen these values for $k$ and number of input qubits to improve readability of the circuits. In addition, we also show a simulation of multiples of 14 with 5 qubits using one and two repetitions of the Grover iterator to showcase the difference in the amplified amplitude in both cases.

We have used Qiskit \cite{Qiskit} to generate the circuits and simulate them. To be able to amplify the marked quantum states we have chosen a full superposition of $0s$ and $1s$ as our input state and have applied the Grover's algorithm diffuser \cite{nielsen2002quantum} after the oracle . All the simulations are conducted with 20,000 shots \footnote{Each shot is one simulation of the circuit, the final result is the aggregation of all shots.} as it is the maximum allowed by Qiskit.

 \subsection{Multiples of 3}

\begin{figure}
    \centering
    \includegraphics[width=\textwidth]{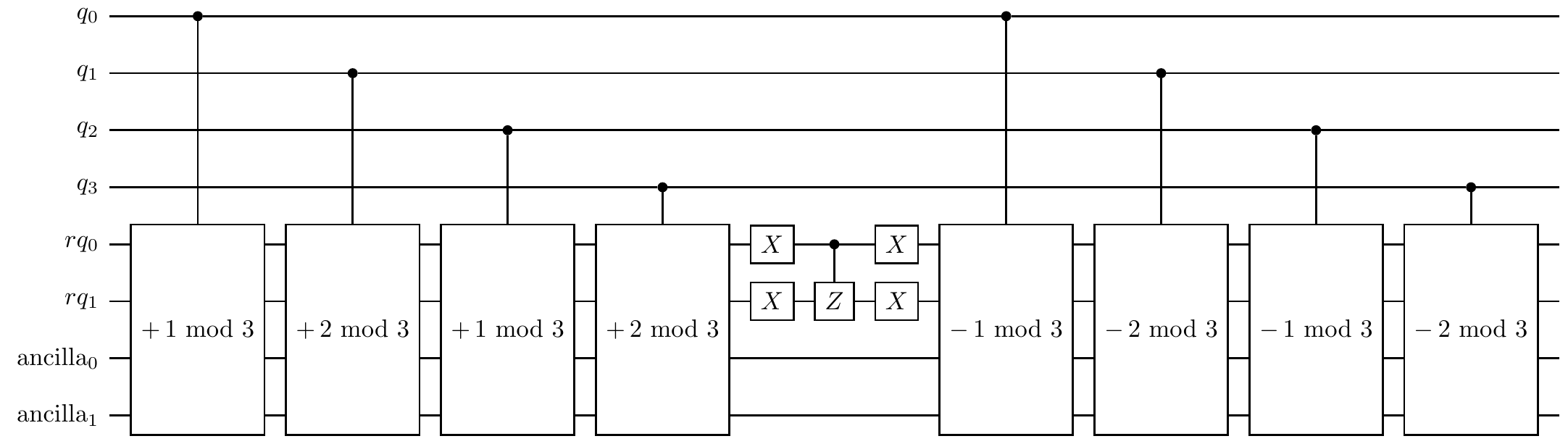}
    \caption{Multiples of 3 oracle with a 4 qubits input register.}
    \label{fig:circuit-multiples-3-4qubits}
\end{figure}

 The `multiples of 3' oracle with 4 qubits as input can be found in figure \ref{fig:circuit-multiples-3-4qubits}. It can be seen that the remainders calculated are 1 and 2, after that the cycle repeats again.
 
\begin{figure}
    \centering
    \includegraphics[width=\textwidth]{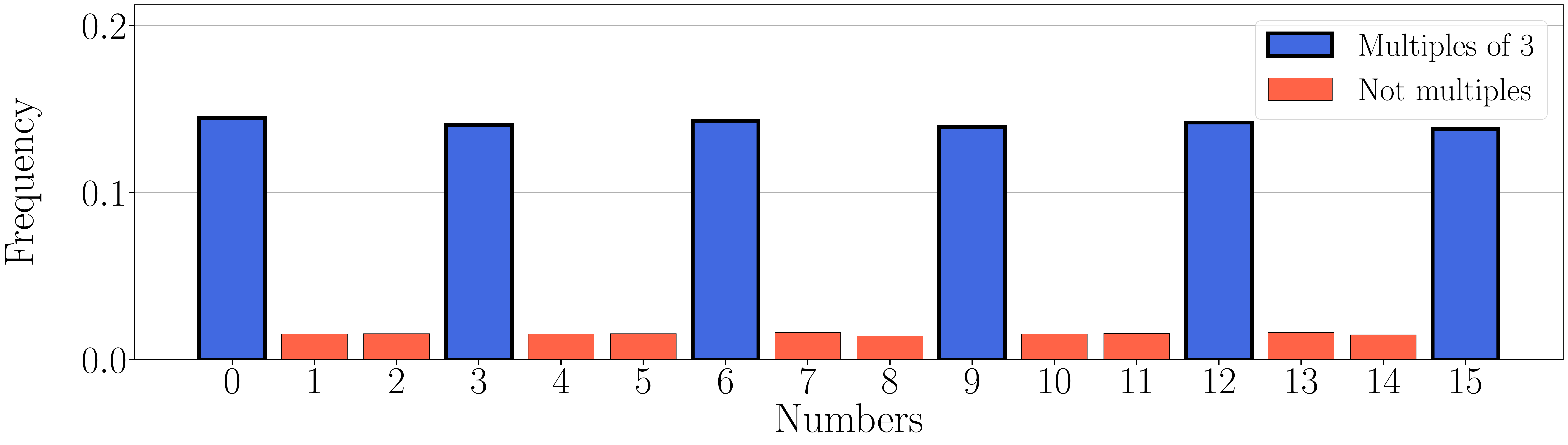}
    \caption{Results of simulating the circuit of multiples of 3 with a 4 qubits input.}
    \label{fig:simulation-multiples-3-4qubits}
\end{figure}

The result of simulating the full circuit, including the initialisation of the state, the shown oracle and the diffuser; are shown on figure \ref{fig:simulation-multiples-3-4qubits}. On the x-axis the final quantum states, on the y-axis the relative frequency. In blue with a thick border, desired states (multiples of 3), in red without border, undesired states (not multiples of 3). It can be noticed that with just one repetition, the desired states are clearly amplified to differentiate the multiples of 3 from the rest of the numbers.

 \subsection{Multiples of 5}
\begin{figure}
    \centering
    \includegraphics[width=\textwidth]{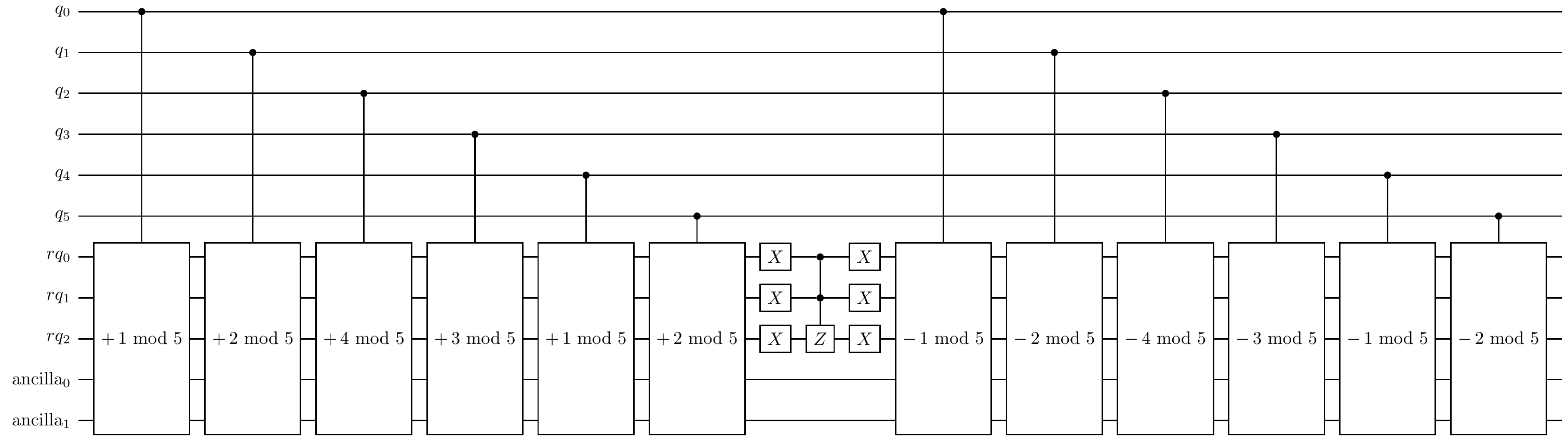}
    \caption{Multiples of 5 oracle with a 6 qubits input register.}
    \label{fig:circuit-multiples-5-6qubits}
\end{figure}

 The `multiples of 5' oracle with 6 qubits as input can be found in figure \ref{fig:circuit-multiples-5-6qubits}. The remainders of the powers of 2 follow the cycle 1, 2, 4, 3. As there are 6 input qubits, the remainders 1 and 2 are repeated.

\begin{figure}
    \centering
    \includegraphics[width=\textwidth]{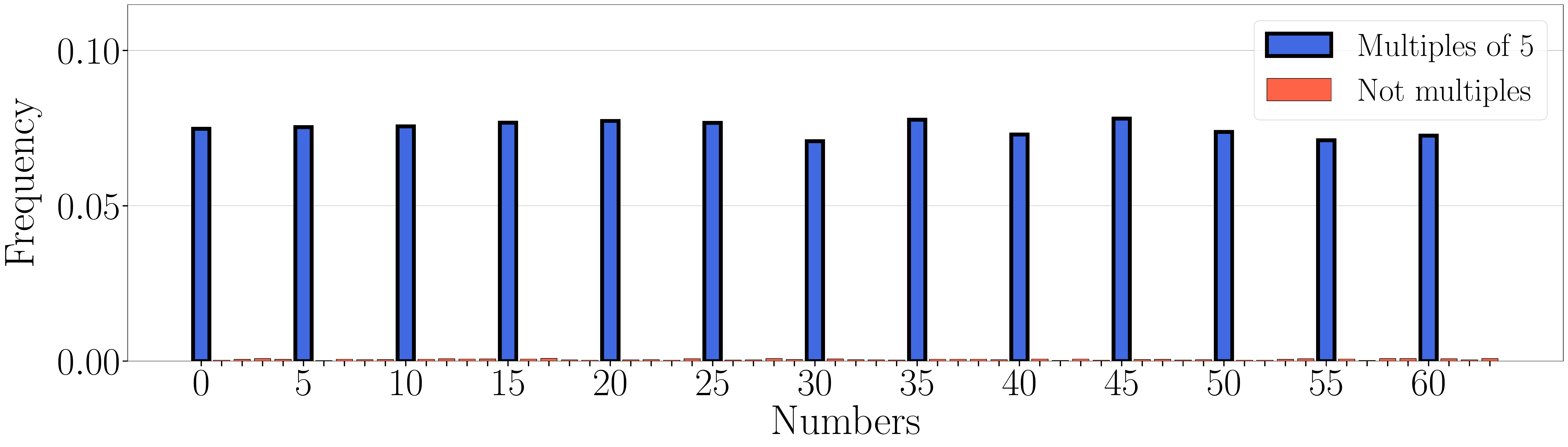}
    \caption{Results of simulating the circuit of multiples of 5 with a 6 qubits input.}
    \label{fig:simulation-multiples-5-6qubits}
\end{figure}

The result of simulating the full circuit, including the initialisation of the state, the shown oracle and the diffuser; are shown on figure \ref{fig:simulation-multiples-5-6qubits}. It can be seen that, in this case, the amplification with one iteration is almost perfect.

\subsection{Multiples of 14}
In this section we show the results of simulating the `multiples of 14' with 5 input qubits in full superposition with one and two repetitions of the Grover iterator. The remainders of the powers of 2 are 1, 2, 4, 8, and 2. We do not show the circuits for the shake of readability, however they can be found in the provided \hyperref[sec:repository]{repository}.

\begin{figure}
    \centering
    \includegraphics[width=\textwidth]{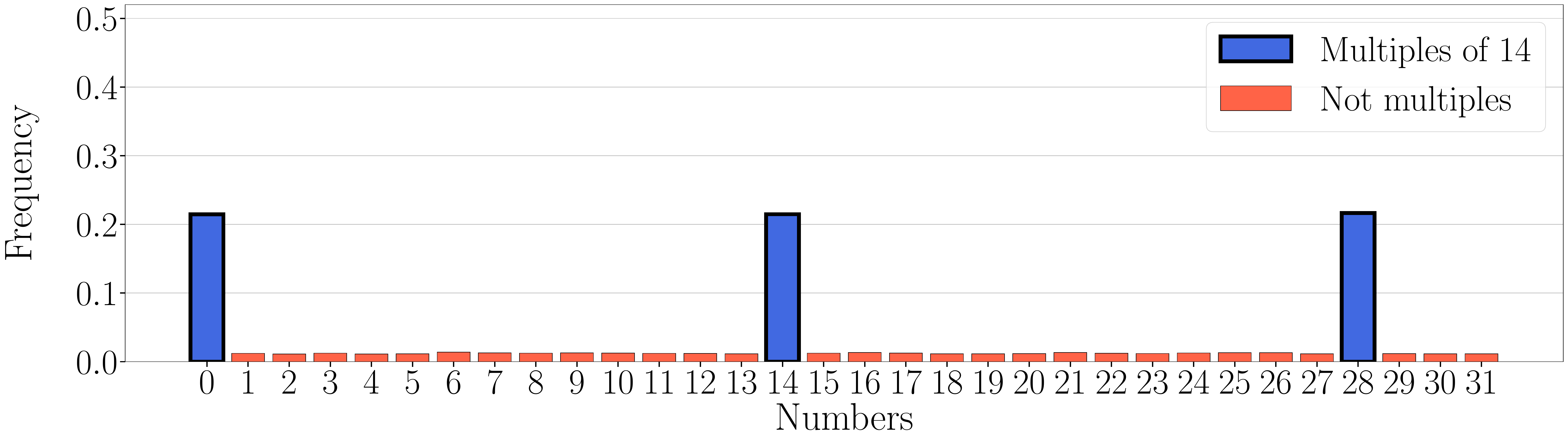}
    \caption{Results of simulating the circuit of multiples of 14 with a 5 qubits input with one repetition of the Grover iterator.}
    \label{fig:simulation-multiples-14-5qubits-onerep}
\end{figure}

Figure \ref{fig:simulation-multiples-14-5qubits-onerep} shows the results of the simulation using one repetition. The total amount of amplification of desired states is $\approx 64\%$. Although from an absolute perspective this may not seem a favourable result, the three desired states (0, 14 and 28, multiples of 14 up to 31) have their amplitude enlarged by a factor $\approx 6.82$. This amplification allows the distinction of multiples of 14 among the input states.

\begin{figure}
    \centering
    \includegraphics[width=\textwidth]{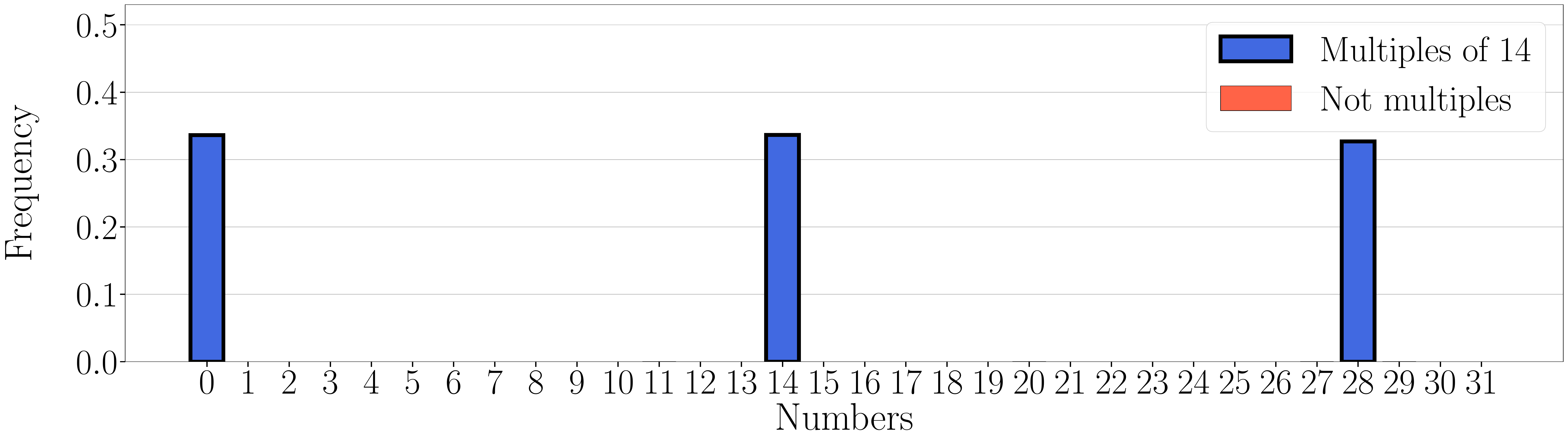}
    \caption{Results of simulating the circuit of multiples of 14 with a 5 qubits input with two repetitios of the Grover iterator.}
    \label{fig:simulation-multiples-14-5qubits-tworeps}
\end{figure}

On figure \ref{fig:simulation-multiples-14-5qubits-tworeps} are depicted the results of the simulation using two repetitions. In this case, the total amount of amplification is $\approx 100\%$. This is the best possible amplification and shows that this oracle may improve its applicability by repeating the pair oracle-diffuser. However, the increased depth of this operation has to be taken into account when implementing the operation in a real quantum device.

\section{Discussion} \label{sec:complexity}
In this section we present both an analysis of the classical calculations required to build the quantum oracle and also a theoretical and empirical study of the depth of the resulting quantum oracle.

Let us bear in mind that $k\in \mathbb{N}$ is the number whose multiples want to be calculated; $n$ is the number of input qubits, in which the multiples of $k$ are going to be calculated; $N=2^n$ is the total number of quantum states; and $n_k$ is the required number of qubits to store the remainders of dividing by $k$ (at most $k-1$).

\subsection{Classical Calculations Complexity}\label{subsec:classical_calculations_complexity}
In this subsection we analyse the classical calculations needed to implement the `multiples of' quantum oracle. This classical part is divided in two tasks. First one is computing the remainders of the powers of 2 divided by $k$. Second task is building the quantum circuit.

The first task consists on the calculations of the remainders \mbox{$r_i \equiv 2^i$ mod $k$,} $0\leq r_i < k$ and $i\in [0, n-1]$. At most, only $n$ remainders need to be calculated, as only $n$ modulo additions are conducted. Therefore, this operation is \mbox{$\mathcal{O}(n) = \mathcal{O}(\log N)$}. The algorithm to do these computations can be found in \hyperref[app:A]{Appendix A}.

The second task is building the quantum circuit. The construction of the controlled circuit `+$r$ mod $k$' which performs the modulo addition is linear on the number of qubits \cite{Shor23}. In this case, the number of qubits on the remainders register, $n_k$. This means that the complexity of this operation is \mbox{$\mathcal{O}(\log k) = \mathcal{O}(n_k)$}. This is smaller than $\mathcal{O}(n)$, otherwise, $k$ would be greater than $N=2^n$ and there would be only one multiple in those integers, the number 0. 

Moreover, the complexity of appending the modulo addition circuits to the full quantum circuit is linear on the number of qubits on the input register, $n$, thus, $\mathcal{O}(n) = \mathcal{O}(\log N)$. The rest of needed appends (Hadamard, $X$ and multi-controlled $Z$) are also linear with the number of qubits. Therefore, the complete procedure required in classical computations holds a complexity of \mbox{$\mathcal{O}(n)=\mathcal{O}(\log N)$}.

It can be noted that for obtaining the multiples of a given number $k$ up to $N$ classically, it is needed to calculate $\lceil N/k\rceil$ multiples, hence, as $k$ is already fixed, this calculation grows exponentially with the number of qubits, \mbox{$\mathcal{O} (N)=\mathcal{O}(2^{n})$}. As discussed above, the complexity of our method is linear on the number of qubits, hence, our method presents an exponential reduction of this complexity.

\subsection{Theoretical Analysis of Quantum Circuit Depth}\label{subsec:theoretical-analysis-quantum-circuit}
As stated in section \ref{sub:algorithm-oracle}, the quantum circuit consists of three registers of qubits, the input qubits, which hold the information for all the possible numbers, formed by $n$ qubits, input from the user. The register which holds the remainder of the numbers, which has $n_k = \lceil \log_2 (k-1) \rceil$ qubits. At most, the remainder of dividing by $k$ is $k-1$, hence not more qubits are required. Finally, an ancilla register with two qubits is needed to perform the modulo addition, as described in detail in \cite{Shor23}. The depth of this circuit is determined by the depth of its two reused oracles, the modulo addition and the phase-marking operation.

The modulo addition `+$r$ mod $k$' has linear depth on the number of qubits, in this case $\mathcal{O}(n_k) = \mathcal{O}(\log k)$, as it is applied on the remainders register. Once $k$ is chosen, the depth of this circuit is fixed. This operation needs to be applied $2n$ times, firstly to compute the remainders and afterwards to uncompute them. Therefore, the depth of this operation is $\mathcal{O}(n) = \mathcal{O}(\log N)$.

The phase-marking operation requires a multi-controlled $Z$-gate. This is implemented following \cite{multicontrol2022}, which provides a linear depth on the number of qubits, $\mathcal{O}(n_k) = \mathcal{O}(\log k)$. As stated in the previous subsection \ref{subsec:classical_calculations_complexity}, this is upper-bounded by $\mathcal{O}(n)$.

Therefore, the depth of the full implementation of the `multiples of' oracle is linear on the number of input qubits $n$, $\mathcal{O}(n)=\mathcal{O}(\log N)$.

\subsection{Empirical Measurement of Circuit Depth}

To further reinforce the depth complexity study, an empirical analysis is also presented. In order to do this, we have generated the oracles for different numbers of $k$, $n_k$, and $n$. To properly perform this analysis, before measuring depth, all the circuits have been transpiled using one of the IBM quamtum computer backends. In particular, the one used has been \textit{fake\_washington\_v2}, which has the same properties (gate set, coupling map, etc) as the real one.

\begin{figure}
    \centering
    \includegraphics[width=\textwidth]{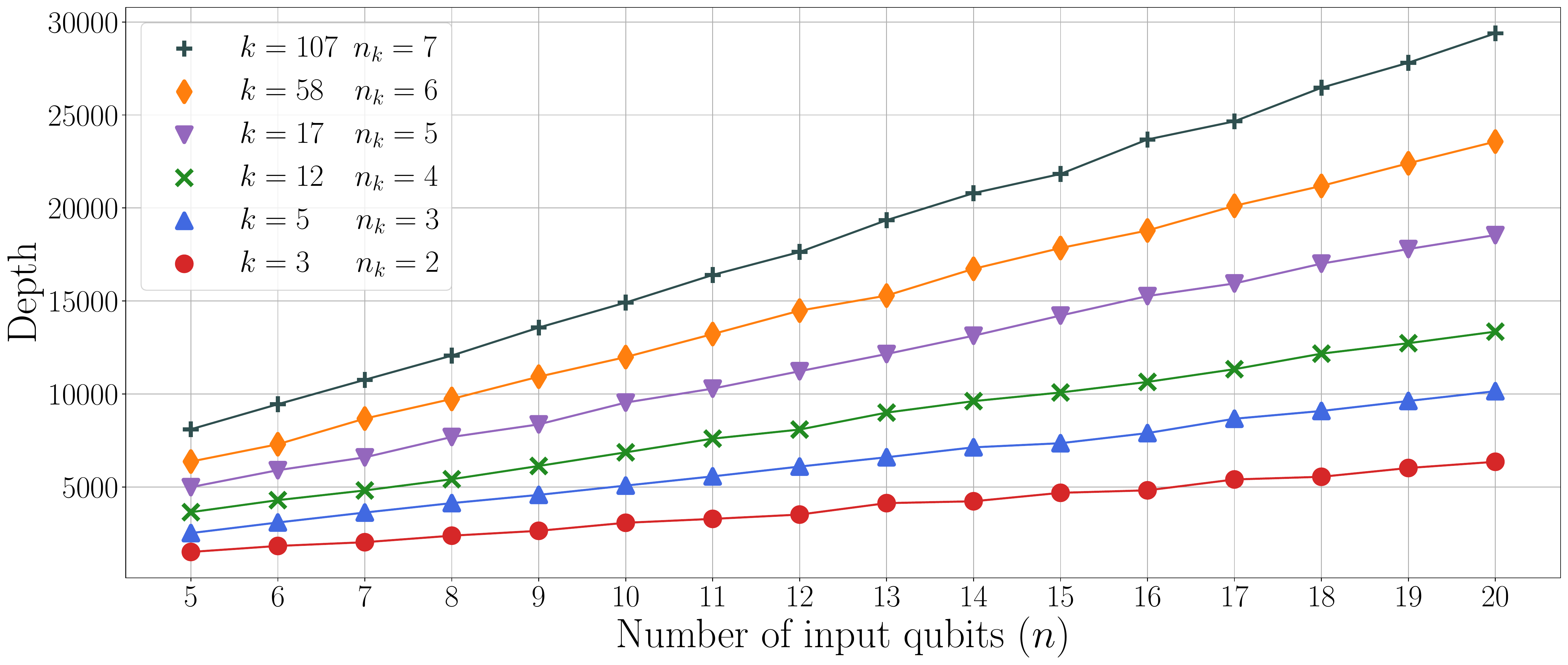}
    \caption{Depth (y-axis) against number of input qubits $n$ (x-axis) for different values of $k$.}
    \label{fig:depths_k}
\end{figure}

Figure \ref{fig:depths_k} shows the depth of the oracle with respect to the number of input qubits $n$, for different values of $k$ and $n_k$. It can be noticed that the depth grows linearly as the number of input qubits increases. This is an expected behaviour as theoretically explained above.

\begin{figure}
    \centering
    \includegraphics[width=\textwidth]{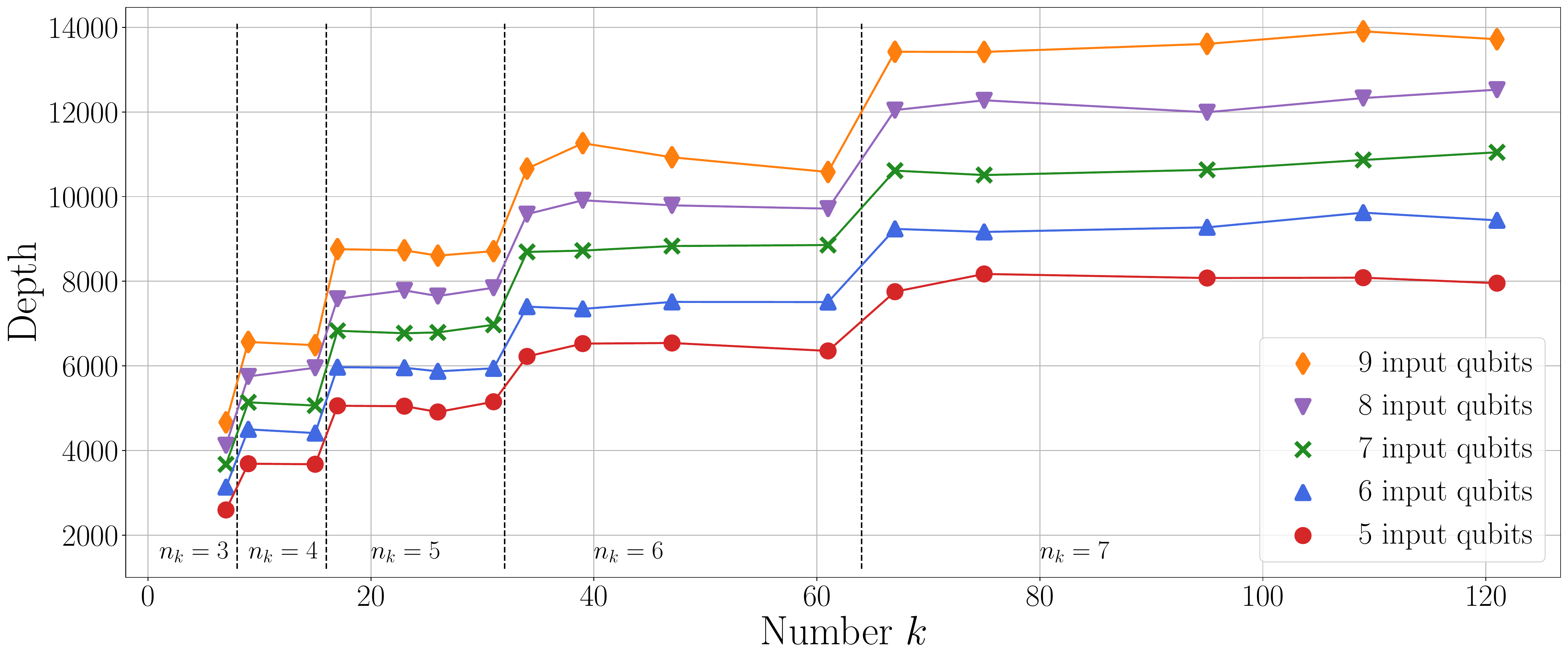}
    \caption{Depth (y-axis) against $k$ (x-axis) for different amounts of input qubits ($n$). The corresponding value of $n_k$ for each $k$ is displayed.}
    \label{fig:depths_qubits_input}
\end{figure}

It can also be noticed that the slope of the graphic grows as the number of qubits in the remainders register $n_k$ increases. This is also an expected behaviour, as the depth of the modulo addition grows linearly with the number of qubits on which it is applied. This behaviour can be observed in figure \ref{fig:depths_qubits_input}.  This figure shows the growth of the depth with respect to the number $k$ whose multiples are to be computed. This analysis has been conducted by choosing several pseudo-random numbers in each interval $[2^{n_k-1}, 2^{n_k})$, with $n_k \in \{3, 4, 5, 6, 7\}$. These intervals are delimited by vertical dotted lines on the figure. It can be observed that the depth for each value holds mostly constant in these intervals. This is, the depth increments are mainly caused by the growing number of $n_k$ qubits required to store the number $k$.

Lines in both figures are mere visual guides and do not represent any data.

\section{Composability and Further Uses}\label{sec:composability}

In this section we show how the proposed oracle can be further reused by providing some examples. First, we showcase how the `multiples of' oracle can be composed with other oracles.  Second, we explain how the oracle can be modified to obtain, instead of multiples of a number $k$, numbers with a determined remainder when dividing by $k$. Last, both of these options are combined. Both circuit and results of simulations are displayed in each case. The conditions for the simulations are the same as previously described in section \ref{sec:simulations}.

\subsection{Multiples and Less-than Oracle}\label{subsec:multiples_5_lessthan_14}
We show an example on how to obtain the multiples of a given number $k$ smaller than $m$. In order to do so, the `multiples of' oracle and the `less-than' oracle \cite{sanchez2023automatic} are composed\footnote{The `multiples of' oracle can be combined with any other phase-marking oracle}. However, this composition is not trivial since it must be applied in an specific way. The oracle to compose with (`less-than' in this example) must be applied controlled by the qubits in the remainders register $rq_0, \ldots, rq_{n_k-1}$ and targeted on the input register $q_0, \ldots , q_{n-1}$. This oracle substitutes the multi-controlled $Z$-gate which is used originally to mark all the multiples. 

\begin{figure}
    \centering
    \includegraphics[width=\textwidth]{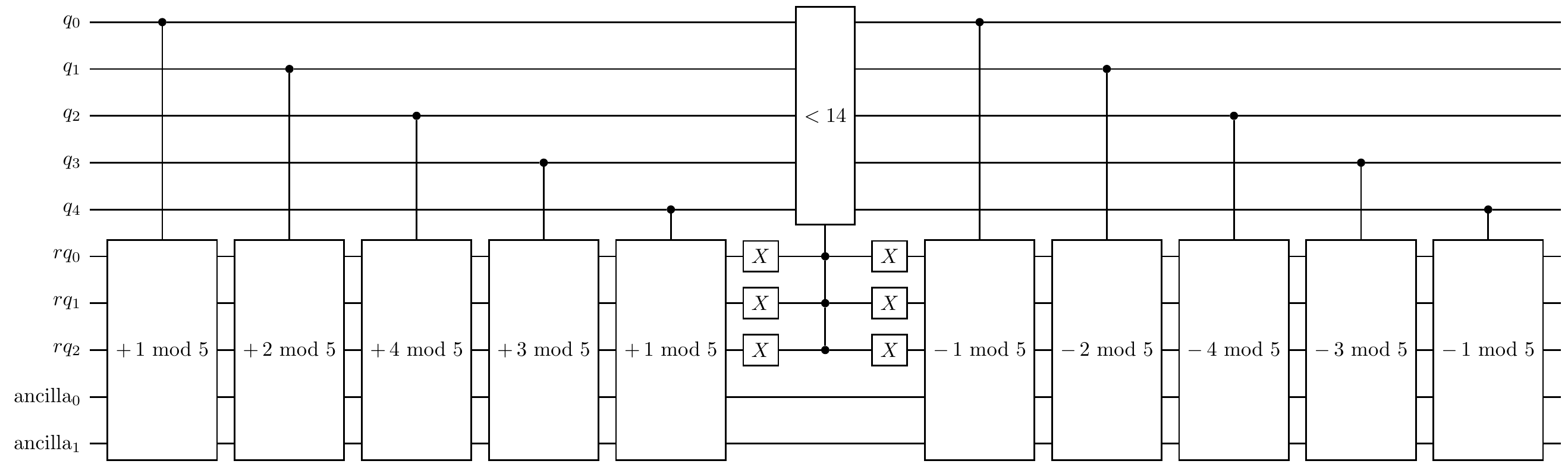}
    \caption{Multiples of 5 oracle combined with less than 14 oracle with a 5 qubits input.}
    \label{fig:circuit_5_K_lessthan_14_w_5_qubits}
\end{figure}

\begin{figure}
    \centering
    \includegraphics[width=\textwidth]{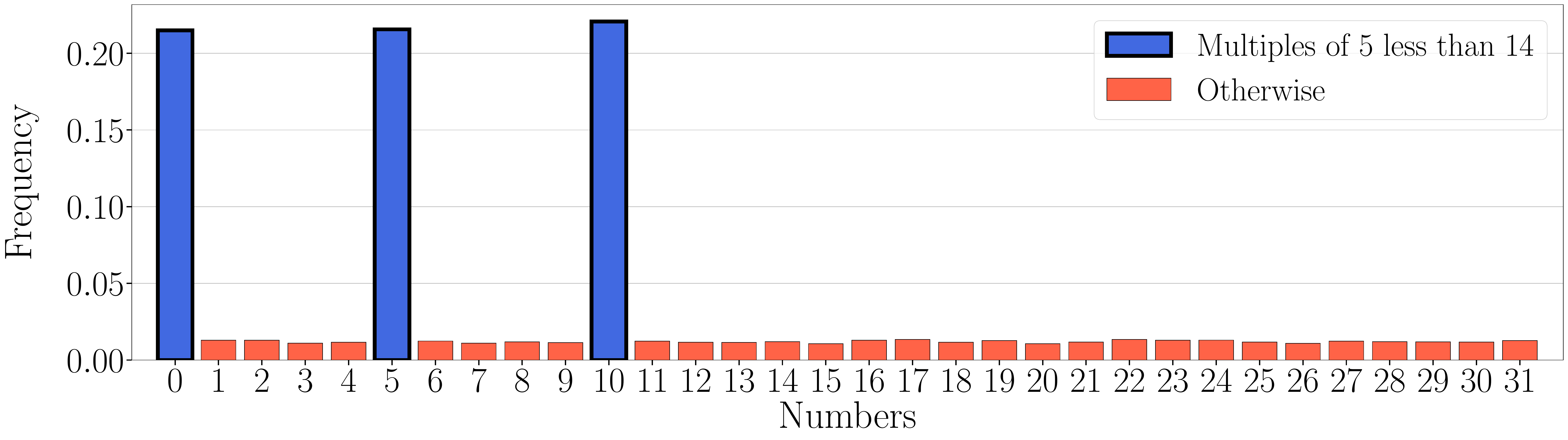}
    \caption{Results of simulating the circuit of multiples of 5 less than 14 with a 5 qubits input.}
    \label{fig:simulation_5_K_lessthan_14_w_5_qubits}
\end{figure}

In this example, the choices are $k=5$, $m=14$, $n=5$. Hence, the desired states are the multiples of 5 smaller than 14 from 0 to 31. The implementation of this oracle can be found in figure \ref{fig:circuit_5_K_lessthan_14_w_5_qubits}. The results of the simulation using only one repetition of the Grover iterator is in figure \ref{fig:simulation_5_K_lessthan_14_w_5_qubits}. The results are, as expected, the states amplified of the multiples of 5 less than 14.

\subsection{Numbers with any Remainder}
\begin{figure}
    \centering
    \includegraphics[width=\textwidth]{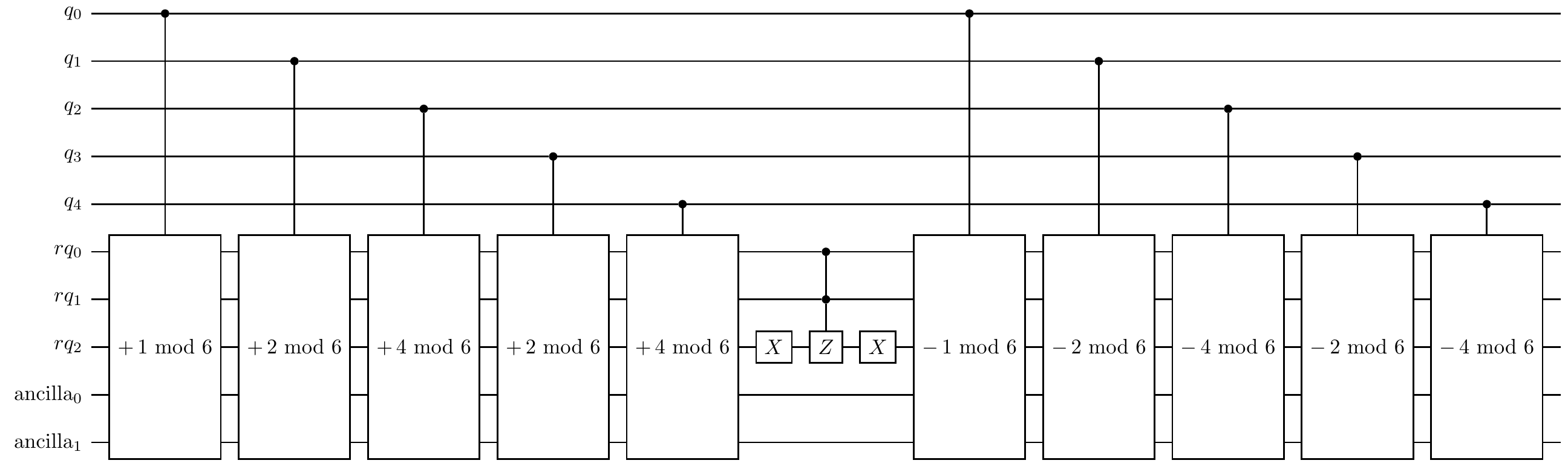}
    \caption{Numbers $p\equiv 3$ mod $6$ oracle with a 5 qubits input.}
    \label{fig:circuit_n_equiv_3_mod_6_w_5_qubits}
\end{figure}

\begin{figure}
    \centering
    \includegraphics[width=\textwidth]{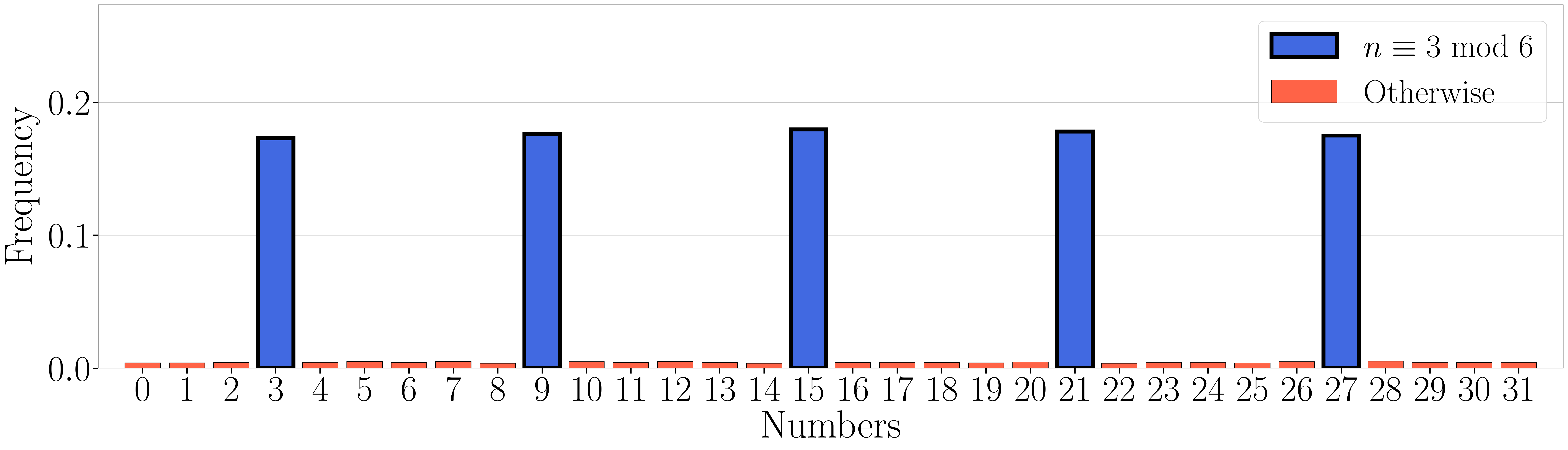}
    \caption{Results of simulating the circuit of numbers $n\equiv 3$ mod $6$ with a 5 qubits input.}
    \label{fig:simulation_n_equiv_3_mod_6_w_5_qubits}
\end{figure}

This subsection shows how to change the multiples oracle in order to obtain numbers with any remainder $r$ when dividing by a given integer $k$. The operation `multiples of' explained so far is the particular case $r=0$. In this example, we show the oracle taking $k=6$, $r=3$, $n=5$, formally, $p\equiv 3$ mod $6$. The oracle can be found in figure \ref{fig:circuit_n_equiv_3_mod_6_w_5_qubits}. Notice that, when giving a $\pi$-phase with gate $CCZ$ in the remainders register, there are only $X$ gates in the qubit $rq_2$, hence marking those states where $|rq_2\,rq_1\,rq_0\rangle = |011\rangle =|3\rangle = |r\rangle$. The results of the simulation using only one repetition of the Grover iterator is shown in figure \ref{fig:simulation_n_equiv_3_mod_6_w_5_qubits} and match the expected results for this operation.

\subsection{Numbers with any Remainder and Range of Integers}
\begin{figure}
    \centering
    \includegraphics[width=\textwidth]{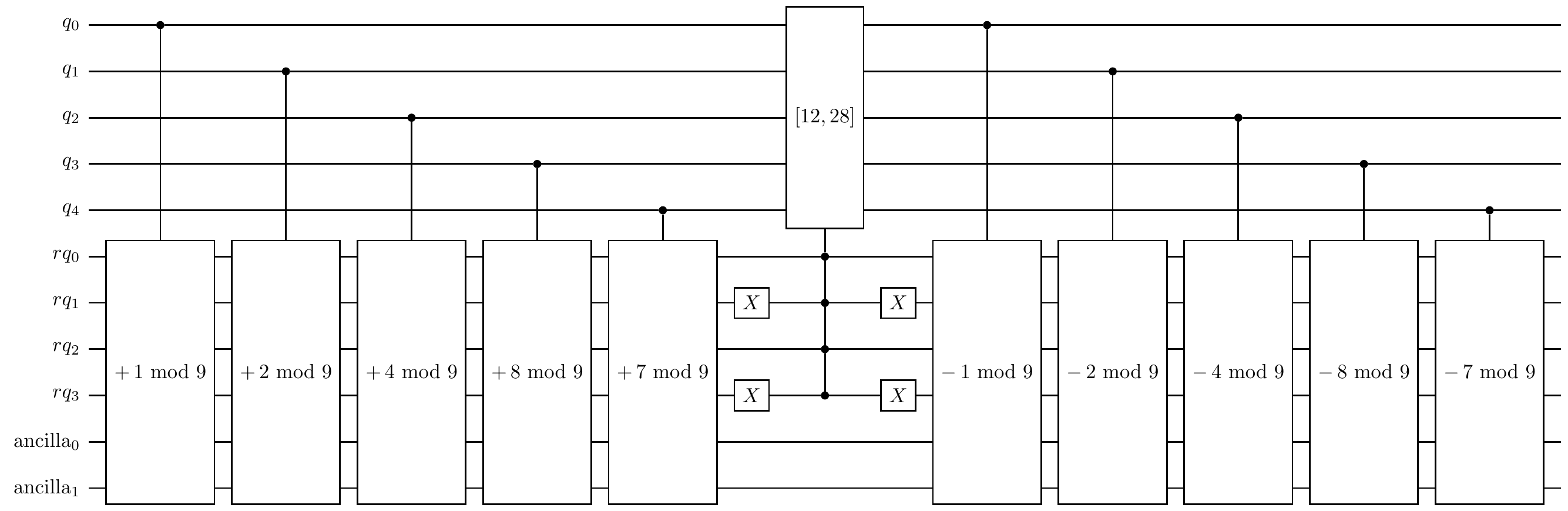}
    \caption{Numbers $n\equiv 5$ mod $9$ with $n\in[12, 28]$ oracle with a 5 qubits input.}
    \label{fig:circuit_n_equiv_5_mod_9_in12_28_w_5_qubit}
\end{figure}

\begin{figure}
    \centering
    \includegraphics[width=\textwidth]{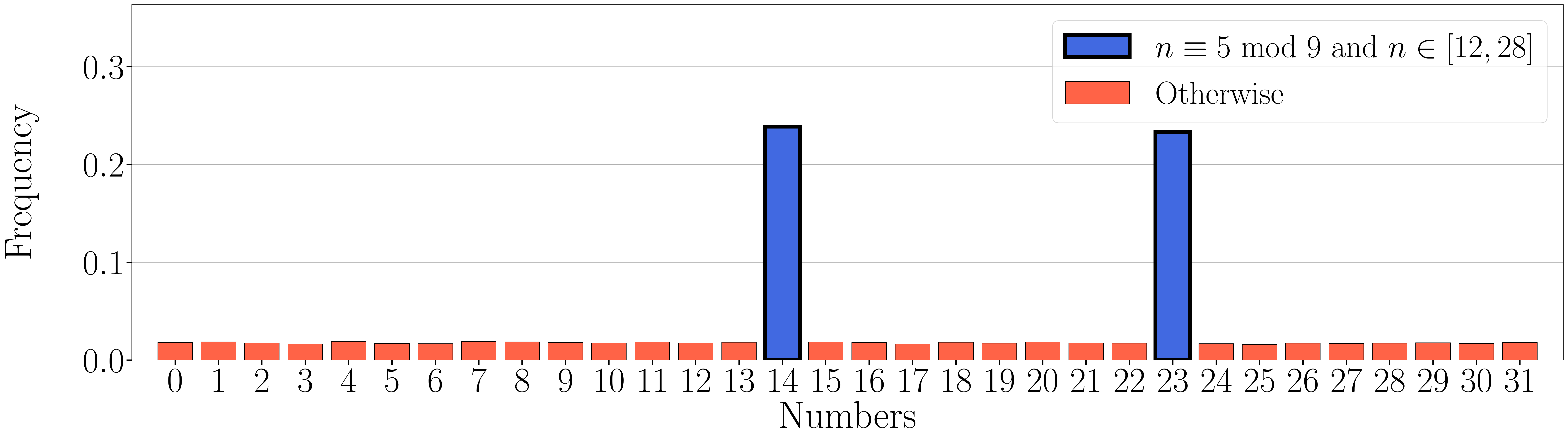}
    \caption{Results of simulating the circuit of numbers $n\equiv 5$ mod $9$ with $n\in[12, 28]$ with a 5 qubits input.}
    \label{fig:simulation_n_equiv_5_mod_9_in12_28_w_5_qubit}
\end{figure}

This subsection shows how to combine the oracle of numbers with a determined remainder when dividing by a number and the range of integers oracle presented in \cite{sanchezrivero2023initial}. For instance, here we show the oracle for integers \mbox{$p\equiv 5$ mod $9$} and \mbox{$p\in[12, 28]$}. The oracle can be found in figure \ref{fig:circuit_n_equiv_5_mod_9_in12_28_w_5_qubit}. Notice that, as in subsection \ref{subsec:multiples_5_lessthan_14}, there is an oracle controlled by the qubits in the remainders register. However, in this case, the $X$ gates are arranged such that the oracle is activated when the the qubits in the remainders register are in the state $|rq_3\,rq_2\,rq_1\,rq_0\rangle = |0101\rangle = |5\rangle$. The results of the simulation using only one repetition of the Grover iterator is shown in figure \ref{fig:simulation_n_equiv_5_mod_9_in12_28_w_5_qubit}.

\section{Conclusions} \label{sec:conclusions}
In this work we have presented a method to build an efficient oracle for phase-marking multiples of a given number. We have shown the theoretical ideas behind this construction and how to build the quantum circuit.  Moreover, we have conducted a theoretical analysis of the complexity of both the classical calculations needed to build the oracle and the oracle itself. The result of this analysis is that our method leads to an exponential speedup over the classical one. Finally, further functionalities are explored. Through examples and simulations we show how to compose the `multiples of' oracle with other oracles and also how numbers with other properties can be obtained.

This work is one of the steps taken to create an efficient set of tools of quantum software for working with integers. We hope these tools can be reused by quantum software developers to create new quantum algorithms.

\section*{Acknowledgements}
This work has been financially supported by the Ministry of Economic Affairs and Digital Transformation of the Spanish Government through the QUANTUM ENIA project call - Quantum Spain project, by the Spanish Ministry of Science and Innovation under project PID2021-124054OB-C31, by the Regional Ministry of Economy, Science and Digital Agenda, and the Department of Economy and Infrastructure of the Government of Extremadura under project GR21133, and by the European Union through the Recovery, Transformation and Resilience Plan - NextGenerationEU within the framework of the Digital Spain 2026 Agenda.

We are grateful to COMPUTAEX Foundation for allowing us to use the supercomputing facilities (LUSITANIA II) for calculations.

\section*{Repository} \label{sec:repository}
The code used for this paper can be found in the following repository: \\ \href{https://github.com/JSRivero/oracle-multiples}{https://github.com/JSRivero/oracle-multiples}.

%
%
%

\bibliographystyle{splncs04}
\bibliography{mybib}

\begin{thebibliography}{10}
\providecommand{\url}[1]{\texttt{#1}}
\providecommand{\urlprefix}{URL }
\providecommand{\doi}[1]{https://doi.org/#1}

\bibitem{Shor23}
Beauregard, S.: Circuit for shor's algorithm using 2n+3 qubits  (2002).
  \doi{10.48550/ARXIV.QUANT-PH/0205095},
  \url{https://arxiv.org/abs/quant-ph/0205095}

\bibitem{AA}
Brassard, G., Hoyer, P., Mosca, M., Tapp, A.: Quantum amplitude amplification
  and estimation. Contemporary Mathematics  \textbf{305},  53--74 (2002)

\bibitem{bigOnotation}
Chivers, I., Sleightholme, J., Chivers, I., Sleightholme, J.: An introduction
  to algorithms and the big o notation. Introduction to Programming with
  Fortran: With Coverage of Fortran 90, 95, 2003, 2008 and 77 pp. 359--364
  (2015)

\bibitem{QFTcoppersmith}
Coppersmith, D.: An approximate fourier transform useful in quantum factoring.
  arXiv preprint quant-ph/0201067  (2002)

\bibitem{draper}
Draper, T.G.: Addition on a quantum computer (2000).
  \doi{10.48550/ARXIV.QUANT-PH/0008033},
  \url{https://arxiv.org/abs/quant-ph/0008033}

\bibitem{grover1}
Grover, L.K.: A fast quantum mechanical algorithm for database search (1996).
  \doi{10.48550/ARXIV.QUANT-PH/9605043},
  \url{https://arxiv.org/abs/quant-ph/9605043}

\bibitem{Grover_1998}
Grover, L.K.: Quantum computers can search rapidly by using almost any
  transformation. Physical Review Letters  \textbf{80}(19),  4329--4332 (may
  1998). \doi{10.1103/physrevlett.80.4329}

\bibitem{history}
Hidary, J.D., Hidary, J.D.: A brief history of quantum computing. Quantum
  Computing: An Applied Approach pp. 15--21 (2021)

\bibitem{quantikz}
Kay, A.: Tutorial on the quantikz package. arXiv preprint arXiv:1809.03842
  (2018)

\bibitem{klappenecker2003quantum}
Klappenecker, A., Roetteler, M.: Quantum software reusability. International
  Journal of Foundations of Computer Science  \textbf{14}(05),  777--796 (2003)

\bibitem{Leymann_QuantumAlgorithms}
Leymann, F.: Towards a pattern language for quantum algorithms. In: Quantum
  Technology and Optimization Problems: First International Workshop, QTOP
  2019, Munich, Germany, March 18, 2019, Proceedings 1. pp. 218--230. Springer
  (2019)

\bibitem{montanaro2016quantum}
Montanaro, A.: Quantum algorithms: an overview. npj Quantum Information
  \textbf{2}(1), ~1--8 (2016)

\bibitem{national2019quantum}
{National Academies of Sciences, Engineering, and Medicine and others}: Quantum
  computing: progress and prospects  (2019)

\bibitem{nielsen2002quantum}
Nielsen, M.A., Chuang, I.: Quantum computation and quantum information (2002)

\bibitem{nisq}
Preskill, J.: Quantum computing in the nisq era and beyond. Quantum
  \textbf{2}, ~79 (2018)

\bibitem{sanchez2023automatic}
Sanchez-Rivero, J., Talaván, D., Garcia-Alonso, J., Ruiz-Cortés, A., Murillo,
  J.M.: Automatic generation of an efficient less-than oracle for quantum
  amplitude amplification (2023). \doi{10.48550/ARXIV.2303.07120},
  \url{https://arxiv.org/abs/2303.07120}

\bibitem{sanchezrivero2023initial}
Sanchez-Rivero, J., Talaván, D., Garcia-Alonso, J., Ruiz-Cortés, A., Murillo,
  J.M.: Some initial guidelines for building reusable quantum oracles (2023).
  \doi{10.48550/arXiv.2303.14959}, \url{https://arxiv.org/abs/2303.14959}

\bibitem{multicontrol2022}
da~Silva, A.J., Park, D.K.: Linear-depth quantum circuits for multiqubit
  controlled gates. Phys. Rev. A  \textbf{106},  042602 (Oct 2022).
  \doi{10.1103/PhysRevA.106.042602},
  \url{https://link.aps.org/doi/10.1103/PhysRevA.106.042602}

\bibitem{Qiskit}
tA~v, A., ANIS, M.S., Abby-Mitchell, Abraham, H., AduOffei, Agarwal, R.,
  Agliardi, G., Aharoni, M., Ajith, V., et~al.: Qiskit: An open-source
  framework for quantum computing (2021). \doi{10.5281/zenodo.2573505}

\bibitem{Zhao}
Zhao, J.: Quantum software engineering: Landscapes and horizons (2021).
  \doi{10.48550/ARXIV.2007.07047}, \url{https://arxiv.org/abs/2007.07047}

\end{thebibliography}
\newpage
\section*{Appendix A}\label{app:A}

\SetKwComment{Comment}{/* }{ */}

\begin{algorithm}
\caption{Algorithm for computing the remainders of the first $n$ powers of 2 when divided by $k$}\label{algorithm-list-remainders}
\KwData{Number of powers $n$ and a natural number $k$}
\KwResult{List of remainders $r_i$ of $2^i$ when divided by $k$ \mbox{$r_i\equiv 2^i$ mod $k$ for $i \in [0, n-1]$}}

list\_remainders $\gets list(n)$\;
$r \gets 1$ \Comment*[r]{as $2^0 \equiv 1$ mod $k$ for any $k\in\mathbb{N}$}
\For{$i=1$ to $n-1$}{
    $r' \gets 2\cdot r$\;
    \eIf{$r' < k$}{
        $r\gets r'$
    }{
        $r\gets r' - k$
    }
    list\_remainders[$i$] $\gets r$
}

\end{algorithm}

It can be noticed that this algorithm performs at most 3 operations each iteration, and has $n$ iterations, hence its complexity is $\mathcal{O}(n)$. 

\end{document}